\documentclass
[twocolumn,aps,prd,amsmath,amssymb,floatfix]
{revtex4-1}
\usepackage{silence} 
\usepackage{CJK}%,pinyin}            % chinese fonts
\usepackage[dvips]{graphicx}
\usepackage{bm}                      % bold math
\usepackage{mathptmx}                % math+times fonts
\usepackage{dcolumn}                 % Align table columns on decimal point
\usepackage{multirow}                % tables
\usepackage{xcolor}                  %
\usepackage[
breaklinks,pdfstartview=FitH,CJKbookmarks=true,
bookmarksnumbered=true,bookmarksopen=true,pdfborder={0 0 1},
colorlinks=true,linkcolor=blue,urlcolor=blue,anchorcolor=blue,citecolor=blue]
{hyperref}

\begin{document}

\def\be{\begin{equation}} \def\ee{\end{equation}}
\def\bal#1\eal{\begin{align}#1\end{align}}
\def\ms{M_\odot}
\def\mmax{M_\text{max}}
\def\ra{\rightarrow}
\def\de{\Delta}
\def\fm3{\;\text{fm}^{-3}}
\def\gc3{\,\text{g/cm}^3}
\def\rdu{\rho_\text{DU}} \def\xdu{x_\text{DU}} \def\mdu{M_\text{DU}}
\def\r1s0{\rho_{1S0}} \def\m1s0{M_{1S0}}
\long\def\hj#1{\color{red}#1\color{black}}
\long\def\OFF#1{}

%------------------------------------------------------------------------------

\title{
%Direct Urca processes and nuclear pairing gaps in neutron star cooling}
Neutron star cooling and mass distributions}

%\begin{CJK*}{GB}{gbsn} % Use default fonts from CJK     emacs: raw-text
%\begin{CJK*}{GBK}{Song}
\begin{CJK*}{UTF8}{gbsn}
%\begin{CJK*}{UTF8}{Song}
% iconv -f utf8 -t gb2312 GB_sample.tex > test.tex ; pdflatex test.tex

\author{H. C. Das$^1$}
\author{Jin-Biao Wei (魏金标)$^2$} %jinbiao.wei@ct.infn.it
\author{G. F. Burgio$^1$}
\author{H.-J. Schulze$^1$}

\affiliation{
$^1$INFN Sezione di Catania and Dipartimento di Fisica,
Universit\'a di Catania, Via Santa Sofia 64, 95123 Catania, Italy\\
%\affiliation{
$^2$Physics Department, University of Geosciences, Wuhan, P.R.~China\\
}

\date{\today}

\begin{abstract}
We study the cooling of isolated neutron stars,
employing different nuclear equations of state
with or without active direct Urca process,
and investigate the interplay with the nuclear pairing gaps.
We find that a consistent description of all current cooling data
requires fast direct Urca cooling and reasonable proton 1S0 gaps,
but no neutron 3P2 pairing.
We then deduce the neutron star mass distributions compatible
with the cooling analysis and compare with current theoretical models.
Reduced 1S0 gaps
and unimodal mass distributions
are preferred by the analysis.
The importance of statistical and systematic errors is also investigated.
\end{abstract}

%\vskip1cm
\maketitle
\end{CJK*}

%\keyword{neutron star; nuclear superfluidity; nuclear equation of state}

%==============================================================================
\section{Introduction}

The cooling properties of neutron stars (NSs),
observationally accessible in terms of
temperature (or luminosity) vs. age relations,
are an important tool to obtain a glimpse
on the internal structure and composition of NS matter
\cite{Yakovlev01,Yakovlev14,Miller21b,Burgio21}.
This information is complementary to global observables like
gravitational mass, radius, and tidal deformability,
accessible by other observational methods.

Together, these combined data of a single NS ideally
would be able to constrain the relevant equation of state (EOS) of NS matter.
Currently, due to the scarcity of such combined data,
and also the ambiguities and uncertainties of theoretical models
for the EOS, this goal has not been achieved.
However, recently great progress has been made
regarding observational information,
and cooling data are now available for about 60 objects
\cite{Potekhin20,Cooldat},
while global observables have been strongly constrained
by gravitational-wave observations in particular
\cite{Abbott18,Radice18,Raaijmakers21,Burgio21}.
This has allowed to further restrict the currently `valid'
theoretical EOSs \cite{Wei20c,Burgio21}.

This article is an attempt to update the cooling calculations
to the new data available,
both regarding cooling curves and theoretical EOS,
following our previous articles on this topic
\cite{Taranto16,Fortin18,Wei19,Wei20,Wei20b}.
In particular,
(a) it has become increasingly clear
\cite{Beznogov15,Beznogov15b,Potekhin20,Leinson22,Burgio21}
that fast neutrino cooling processes are required in order to explain
cold and not too old objects,
while also sufficiently slow cooling must be accommodated theoretically
to cover old and warm objects;
(b) the permissible nuclear EOS is more constrained now in terms of
maximum mass, radius, and tidal deformability,
so that several EOSs used in the past for cooling calculations
are not suitable any more.

Another particular feature of our work is the fact that now the
number of available cooling data is becoming sufficiently large
to allow a combined analysis of cooling properties
and NS mass distributions,
as initiated in \cite{Popov06,Beznogov15,Beznogov15b,Wei19,Wei20}.
This permits in particular to draw conclusions regarding the
superfluid properties of NS matter,
and is an important objective of the present article as well.

This paper is organized as follows.
In Sec.~\ref{s:eos} we give a brief overview of the theoretical framework,
regarding the nuclear EOSs,
the various cooling processes,
and the related nucleonic pairing gaps.
Sec.~\ref{s:res} is devoted to the presentation and discussion
of the cooling diagrams and their dependence on the various
theoretical degrees of freedom.
The relation between cooling diagrams and NS mass distributions
is exploited in Sec.~\ref{s:cor}
to obtain information on the pairing gaps of the matter.
In Sec.~\ref{s:err} we investigate the effect of the important
statistical and systematic errors.
Conclusions are drawn in Sec.~\ref{s:end}.

%==============================================================================
\section{Formalism}
%\section{The nuclear equation of state}
\label{s:eos}

\begin{table*}[t]%.............................................................
\renewcommand{\arraystretch}{1.1}
\caption{
Saturation properties of the selected EOSs, i.e.,
density $\rho_0$,
binding energy per nucleon $E_0$,
compressibility $K_0$,
symmetry energy $S_0$,
and its derivative $L$.
Also given are the DU onset proton fraction $\xdu$ and density $\rdu$.
Structure properties of spherically-symmetric NSs, i.e.,
maximum mass $\mmax$,
tidal deformability $\Lambda_{1.4}$,
and radius $R_{1.4}$ of a $1.4\ms$ NS
are also listed.}
\label{t:eos}
%\medskip
\begin{ruledtabular}
\begin{tabular}{lcccccccccc}
  EOS & $\rho_0$ & $-E_0$ & $K_0$ & $S_0$ & $L$ & $\xdu$ & $\rdu$ & $\mmax$ & $\Lambda_{1.4}$ & $R_{1.4}$\\
  & ($\fm3$) & (MeV) & (MeV) & (MeV) & (MeV) & & ($\fm3$) & ($\ms$) & & (km)\\
\hline\\[-3mm]
 V18  & 0.178 & 13.9 & 207 & 32.3 & 63 & 0.135 & 0.37 & 2.36 & 440 & 12.3 \\
 DD2  & 0.149 & 16.0 & 243 & 31.7 & 55 &    -- &   -- & 2.42 & 680 & 13.2 \\
 TW99 & 0.153 & 16.2 & 240 & 32.8 & 55 &    -- &   -- & 2.08 & 400 & 12.3 \\
 Exp. & ~0.14--0.17 & 14--16 & 200--260 & 28--35 & 30--90 & & &
 $>2.35\pm0.17$ & 70--580 & 11.8--13.1 \\
 Ref. & \cite{Margueron18} %audi2012,2013ADNDT..99...69A}
      & \cite{Margueron18} %brown,browne}
      & \cite{Shlomo06,Piekarewicz10}
      & \cite{LiHan13,Oertel17} & \cite{LiHan13,Oertel17} & &
      & \cite{Romani22} & \cite{Abbott18} & \cite{Abbott18} \\
\end{tabular}
\end{ruledtabular}
\end{table*}%..................................................................

In this work we employ a purely nucleonic scenario,
where NS matter is composed of nucleons and leptons only.
Exotic components like hyperons or quark matter are not considered
\cite{Burgio21}.
Even in this case,
the solution of the many-body problem for nuclear matter is still
a very challenging theoretical task.
The currently used many-body methods include ab-initio microscopic methods
and phenomenological approaches.
The former ones start from bare two- and three-nucleon interactions
able to reproduce nucleon scattering data
and properties of bound few-nucleon systems,
whereas the latter ones use effective interactions with a simple structure
dependent on a limited number of parameters,
usually fitted to different properties of finite nuclei and nuclear matter.
The ab-initio methods can be employed
only for the description of homogeneous matter in the NS core,
whereas the phenomenological approaches are well suited
also for the clustered matter typical of the NS crust.
A very rich literature does exist on this topic,
and the interested reader is referred to the recent Ref.~\cite{Burgio21}
for details on the current state of the art.

\subsection{Cooling processes}

In the context of NS cooling the one key property of the nuclear EOS
is whether it allows fast direct Urca (DU) cooling
by a large enough proton fraction
\cite{Yakovlev01,Page06,Page06b,Lattimer07,Yakovlev14,Potekhin15,Burgio21}.
The DU process is the most efficient one among all possible cooling reactions
involving nucleons and neutrino emission
and depends on the NS EOS and composition.
It involves a pair of charged weak-current reactions,
\be
 n \ra p + l + \bar{\nu}_l
\quad \text{and} \quad
 p + l \ra n + \nu_l \:,
\label{e:DU}
\ee
being $l=e,\mu$ a lepton and $\nu_l$ the corresponding neutrino.
Those reactions are allowed by energy and momentum conservation \cite{Lattimer91}
only if $k_F^{(n)} < k_F^{(p)} + k_F^{(l)}$,
where $k_F^{(i)}$ is the Fermi momentum of the species $i$.
This implies that the proton fraction should be larger than a threshold value
(about 13\%)
for the DU process to take place,
and therefore the NS central density should be larger
than the corresponding threshold density.

Thus different EOSs predict different DU threshold densities.
Microscopic EOSs tend to predict higher proton fractions than phenomenological
ones \cite{Burgio21},
therefore we choose here a microscopic Brueckner-Hartree-Fock (BHF) EOS
and two phenomenological relativistic-mean-field (RMF) models,
with and without DU cooling, respectively.

We employ the latest version of a BHF EOS
obtained with the Argonne V18 $NN$ potential and compatible three-body forces
\cite{Li08a,Li08b,Liu22,Liu23},
see \cite{Baldo99,Baldo12} for a more detailed account.
This EOS is compatible with all current low-density constraints
\cite{Wei20b,Burgio21,Burgio21b}
and in particular also with those imposed on NS maximum mass
$\mmax>2\ms$
\cite{Antoniadis13,Arzoumanian18,Cromartie20},
radius $R_{1.4}\approx11$--$13\,$km
\cite{Riley21,Miller21,Pang21,Raaijmakers21},
and tidal deformability $\Lambda_{1.4}\approx70$--$580$
\cite{Abbott17,Abbott18,Burgio18,Wei19}.
For the no-DU EOSs we adopt the density-dependent covariant density functionals
TW99 \cite{Typel99} and DD2 \cite{Typel10,Compose},
also fulfilling most mentioned constraints.
In Table~\ref{t:eos} the main saturation properties of the selected EOSs
are reported,
in comparison with experimental/observational data.

For completeness, we remind the reader that both BHF and RMF methods
provide the EOS for homogeneous nuclear matter,
and therefore an EOS for the low-density inhomogeneous crustal part
has to be added.
For that,
we adopt the well-known Negele-Vautherin EOS \cite{Negele73}
for the inner crust in the medium-density regime
($0.001\fm3 < \rho < \rho_t$),
and the ones by Baym-Pethick-Sutherland \cite{Baym71} and
Feynman-Metropolis-Teller \cite{Feynman49} for the outer crust
($\rho < 0.001\fm3$).
By imposing a smooth transition
of pressure and energy density between both branches of the betastable EOS
\cite{Burgio10},
one finds a transition density at about $\rho_t \approx 0.08\fm3$.
In any case the NS maximum mass domain is not affected by the crustal EOS,
with a limited influence on the radius and related deformability
for NSs with canonical mass value
\cite{Burgio10,Baldo14,Fortin16,Tsang19}.

We also remind that, besides the DU process,
other cooling reactions come into play and involve nucleon collisions,
the strongest one being the modified Urca (MU) process,
\be
 n + N \ra p + N + l \! + \bar{\nu}_l
\quad \text{and} \quad
 p + N + l \ra n + N + \nu_l \:,
\label{e:MU}
\ee
where $N$ is a spectator nucleon that ensures momentum conservation.
The nucleon-nucleon bremsstrahlung (BS) reactions,
\be
 N+N \ra N+N + \nu+\bar{\nu} \:,
\ee
with $N$ a nucleon and $\nu$,
$\bar{\nu}$ an (anti)neutrino of any flavor,
are also abundant in NS cores,
and their rate increases with the baryon density,
but they are orders of magnitude less powerful than the DU one,
thus producing a slow cooling \cite{Yakovlev01}.
All those cooling mechanisms can be strongly affected
by the superfluid properties of the stellar matter, i.e.,
critical temperatures and gaps in the different pairing channels.
We will turn to this theoretical issue in Sec.~\ref{s:gap}.

\begin{figure}[t]%.............................................................
%\vspace{-9mm}
\centerline{\includegraphics[scale=0.72]{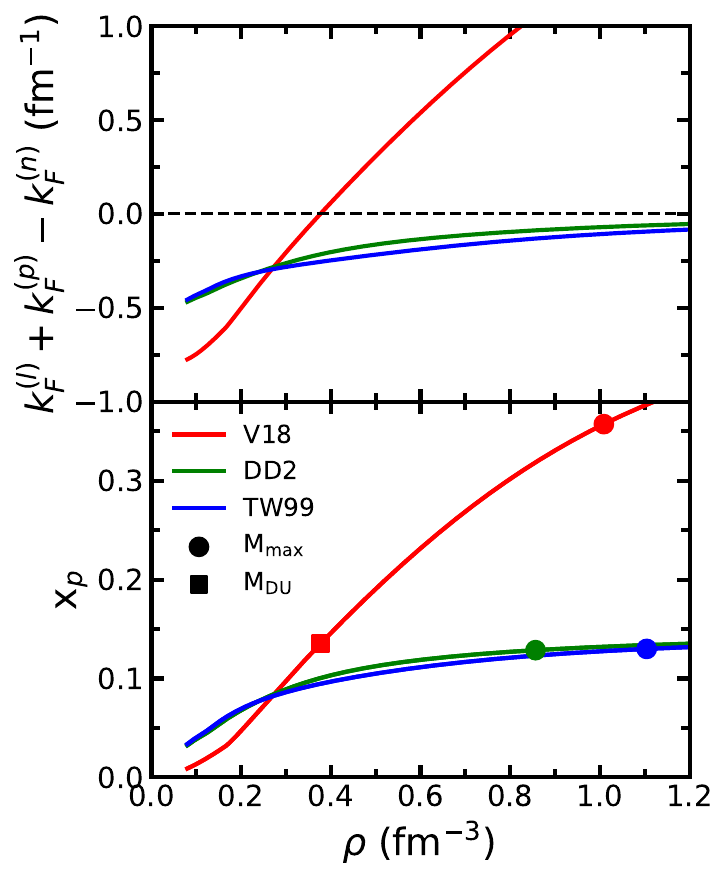}}
\vspace{-4mm}
\caption{
The threshold condition for the DU process (upper panel)
and the proton fraction (lower panel)
vs.~the nucleon density for all chosen EOSs.
Markers indicate the $\mmax$ configurations
and the onset of DU cooling.
}
\label{f:xp}
\end{figure}%..................................................................

%\subsection{Proton fraction}
\subsection{Equation of state} %and stellar structure

In order to illustrate the relevant properties of the chosen EOSs,
we show in Fig.~\ref{f:xp} the DU onset condition (upper panel)
and the proton fraction (lower panel).
One notes that both RMF EOSs predict very similar proton fractions,
not allowing the DU process,
at variance with the BHF V18 EOS,
for which the proton fraction reaches the DU threshold $\xdu=0.135$
at a density $\rdu \approx 0.37\fm3$.
The associated NS mass is $\mdu=1.01\ms$,
hence all NSs described with the V18 EOS can potentially cool very fast.
This is extensively illustrated in the next section.

\begin{figure}[t]%.............................................................
%\vspace{-9mm}
\centerline{\includegraphics[scale=0.72]{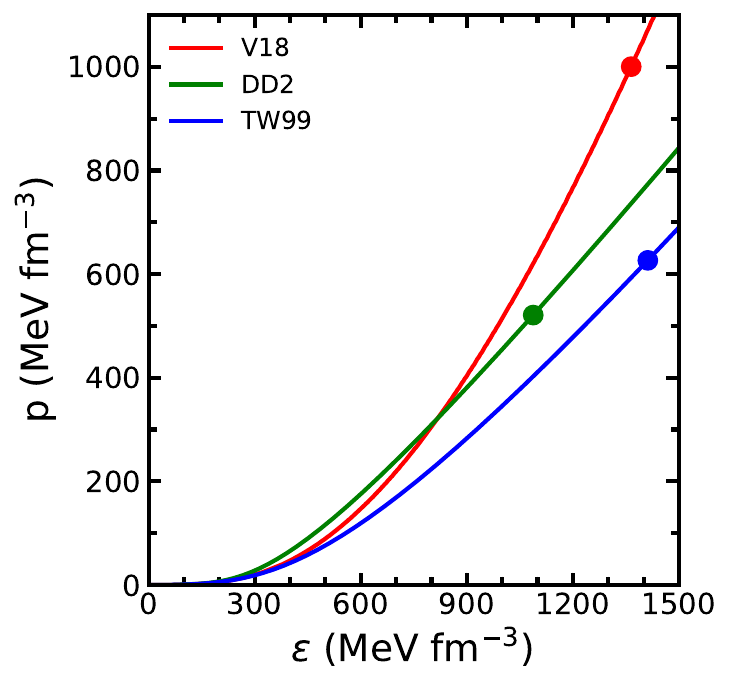}}
\vspace{-4mm}
\caption{
EOSs used in this work.
Markers indicate the $\mmax$ configurations.
}
\label{f:EOS}
\end{figure}%..................................................................

\begin{figure}[t]%.............................................................
%\vspace{-9mm}
\centerline{\includegraphics[scale=0.55]{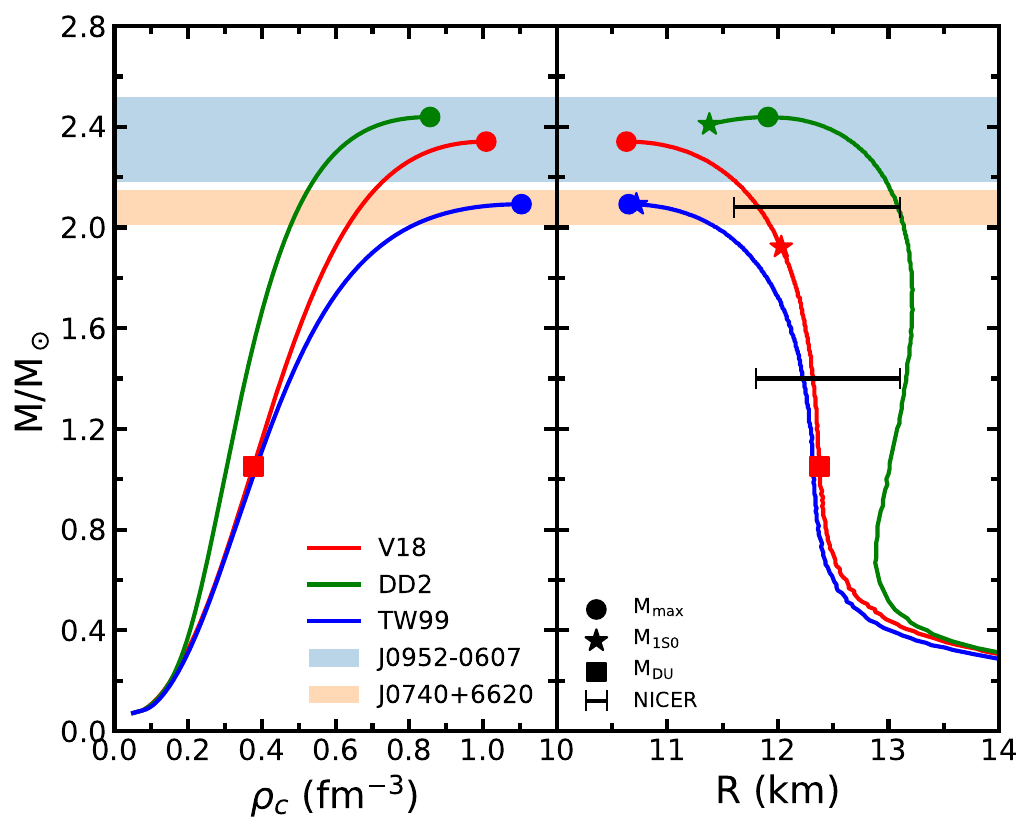}}
\vspace{-3mm}
\caption{
NS gravitational mass vs central baryon density (left)
and radius (right) for the different EOSs.
Dots indicate the configurations of $\mmax$,
and stars those of vanishing p1S0 BCS gap in the NS center.
Observational constraints on masses \cite{Cromartie20,Romani22}
and from NICER \cite{Miller21}
are included.
}
\label{f:MR}
\end{figure}%..................................................................

%\subsection{Equation of state} %and stellar structure

In Fig.~\ref{f:EOS} the final EOS is plotted, i.e.,
pressure vs.~energy density,
after imposing beta-stability and charge neutrality conditions,
whereas in Fig.~\ref{f:MR} we display the resulting
NS mass-radius and mass-central density relations,
obtained by solving the Tolman-Oppenheimer-Volkoff equations.
We stress that the value of the maximum mass for all the selected EOSs
is compatible with the current observational lower limits
\cite{Demorest10,Antoniadis13,Fonseca16,Cromartie20};
the recent data $M=2.35\pm0.17\ms$ for PSR J0952-0607 \cite{Romani22}
is very constraining,
and would actually exclude the TW99 EOS.
We also mention the combined estimates of the mass and radius
of the isolated pulsar PSR J0030+0451 observed recently by NICER,
$M=1.44^{+0.15}_{-0.14}\ms$ and $R=13.02^{+1.24}_{-1.06}\,$km
\cite{Miller19,Riley19}, or
$M=1.36^{+0.15}_{-0.16}\ms$ and $R=12.71^{+1.14}_{-1.19}\,$km
\cite{Miller21,Riley21},
and in particular the result of the combined GW170817+NICER analysis
\cite{Miller21},
$R_{2.08}=12.35\pm0.75\,$km and
$R_{1.4}=12.45\pm0.65\,$km,
with which only the V18 EOS would fully comply,
see Fig.~\ref{f:MR} and Table~\ref{t:eos}.

\OFF{
PSR J1614-2230 ($M = 1.908\pm0.016M_\odot$) \cite{Arzoumanian18},
PSR J0348+0432 ($M = 2.01\pm0.04M_\odot$) \cite{Antoniadis13}, and
PSR J0740+6620  2.14-0.09+0.10 \cite{Cromartie20},
PSR J0740+6620 ($M = 2.08\pm0.07M_\odot$), \cite{Fonseca21},
$M=2.35\pm0.17\ms$ for PSR J0952-0607 \cite{Romani22}

recent mass-radius results of the NICER mission
for the pulsars J0030+0451
\cite{Riley19,Miller19}
$R(1.44^{+0.15}_{-0.14}M_\odot) = 13.02^{+1.24}_{-1.06}$ km
\cite{Riley19} and
$R(1.34^{+0.15}_{-0.16}M_\odot) = 12.71^{+1.14}_{-1.19}$ km
\cite{Miller19},
and for PSR J0740+6620 with
$R(2.08\pm0.07M_\odot) = 13.7^{+2.6}_{-1.5}$ km
\cite{Riley21} and
$R(2.072^{+0.067}_{-0.066}M_\odot) = 12.39^{+1.30}_{-0.98}$ km
\cite{Miller21}.
\cite{Riley21,Miller21,Pang21,Raaijmakers21}.

The combined (strongly model-dependent) analysis
of both pulsars together with GW170817 event observations
\cite{Abbott17,Abbott18}
yields improved limits on
but in particular on the radius $R_{1.4}$, namely
$12.45\pm0.65$ km \cite{Miller21}, %11.8-13.4 (90%) and 12.2-13.1 (68%)
$11.94^{+0.76}_{-0.87}$ km \cite{Pang21}, and
$12.33^{+0.76}_{-0.81}$ km or
$12.18^{+0.56}_{-0.79}$ km \cite{Raaijmakers21}.
}%OFF

%==============================================================================
\subsection{Pairing gaps}
\label{s:gap}

One of the most important nuclear physics input for the NS cooling simulations
are the superfluid properties of stellar matter,
basically the neutron and proton pairing gaps in the different reaction channels
\cite{Yakovlev01,Sedrakian19,Burgio21}.
Usually the most important ones are the proton 1S0 (p1S0)
and neutron 3PF2 (n3P2) pairing channels;
the proton 3PF2 gap is often neglected due to its uncertain properties
at large densities,
while the neutron 1S0 gap in the crust is of little relevance for the cooling.
%Moreover, in our previous works we found that nonzero values of the n3P2 gap
%cannot reproduce the current observational data in our cooling simulations
%\cite{Wei20b}, so we exclude it in the current simulations
%and focus our study on the possible p1S0 gap function $\Delta(\rho)$.

These superfluids are created by the formation of $pp$ and $nn$ Cooper pairs
due to the attractive part of the $NN$ potential,
and are characterized by a critical temperature
$T_c \approx 0.567\Delta$.
The occurrence of pairing when $T < T_c$
leads on one hand to an exponential reduction of the emissivity
of the neutrino processes the paired component is involved in,
and on the other hand to the onset of the ``pair breaking and formation"
(PBF) process with associated neutrino-antineutrino pair emission.
This process starts when the temperature reaches $T_c$
of a given type of baryons,
becomes maximally efficient when $T \approx 0.8\,T_c$,
and then is exponentially suppressed for $T \ll T_c$ \cite{Yakovlev01}.

%Of vital importance for any cooling simulation is the knowledge of the
%1S0 and 3PF2 pairing gaps for neutrons and protons in beta-stable matter,
%which on one hand block the DU and MU reactions,
%and on the other hand open new cooling channels by the PBF mechanism
%for stellar matter in the vicinity of the critical temperature \cite{Yakovlev01}.

In the simplest BCS approximation,
%and detailing the more general case of pairing in the coupled 3PF2 channel,
the relevant pairing gaps
are computed by solving the (angle-averaged) gap equation
\cite{Amundsen85,Baldo92,Takatsuka93,Elgaroy96,Khodel98,Baldo98}
for the $L=0$ and the two-component $L=1,3$ gap functions,
\bal
 \de_0(k) &= - \frac{1}{\pi} \int_0^{\infty}\!\! dk' {k'}^2 \frac{1}{E(k')}
  V_{00}(k,k') \de_0(k') \:,
\label{e:gaps}
\\
 \left(\!\!\!\begin{array}{l} \de_1 \\ \de_3 \end{array}\!\!\!\right)\!(k) &=
 - \frac{1}{\pi} \int_0^{\infty}\!\! dk' {k'}^2 \frac{1}{E(k')}
%\phantom{wwwwwwwww} && \nonumber\\ \times
 \left(\!\!\!\begin{array}{ll}
  V_{11}\!\! & \!\!V_{13} \\ V_{31}\!\! & \!\!V_{33}
 \end{array}\!\!\!\right)\!(k,k')
 \left(\!\!\!\begin{array}{l} \de_1 \\ \de_3 \end{array}\!\!\!\right)\!(k')
%&&
\label{e:gapp}
\eal
with
\bal
 E(k)^2 &= [e(k)-\mu]^2 +
 \de(k)^2 \:, \\ \de(k)^2 &=
 \{ \de_0(k)^2 , \de_1(k)^2 + \de_3(k)^2 \} \:,
\eal
while fixing the (neutron or proton) density,
\be
  \rho = \frac{k_F^3}{3\pi^2}
   = 2 \sum_k \frac{1}{2} \left[ 1 - \frac{e(k)-\mu}{E(k)} \right] \:.
\label{e:rho}
\ee
Here $e(k)=k^2\!/2m$ is the s.p.~energy,
$\mu \approx e(k_F)$ is the chemical potential
determined self-consistently from Eqs.~(\ref{e:gaps}--\ref{e:rho}),
and
\be
   V^{}_{LL'}(k,k') =
   \int_0^\infty \! dr\, r^2\, j_{L'}(k'\!r)\, V^{TS}_{LL'}(r)\, j_L(kr)
\label{e:v}
\ee
are the relevant potential matrix elements
($T=1$ and
$S=1$; $L,L'=1,3$ for the 3PF2 channel,
$S=0$; $L,L'=0$ for the 1S0 channel)
with the bare potential $V$. % = V_{18}$.
The relation between (angle-averaged) pairing gap at zero temperature
%$\de \equiv \sqrt{\de_1^2(k_F)+\de_3^2(k_F)}$
$\de \equiv \de(k_F)$
obtained in this way and the critical temperature of superfluidity is then
$T_c \approx 0.567\de$.

%(If no angle average is performed, the prefactor varies slightly,
%see, e.g.,~\cite{yakov99,Yakovlev01}, % Tab.1 ,potrev15},
%but the angle-average procedure is usually an excellent approximation
%\cite{Baldo92,Papa17}).

At this simplest level of approximation,
the gap is a universal function of the particle density,
$\de_\text{BCS}(\rho)$,
and thus valid for any EOS,
independent of the $NN$ interaction used,
provided that the associated $NN$ phase shifts are reproduced
\cite{Lombardo01,Sedrakian06}. %Lombardo05}.
However, in-medium effects might strongly modify these BCS results,
as both the s.p.~energy $e(k)$ and the interaction kernel $V$ itself
might include the effects of three-body forces and polarization corrections.
It turns out that in the p1S0 channel all these corrections lead
to a suppression of both magnitude and density domain of the BCS gap,
see, e.g, \cite{Zhou04,Baldo07} in the BHF context,
or \cite{Lombardo01,Sedrakian19,Burgio21}
for a collection of different theoretical approaches.

The situation is much worse for the gap in the n3P2 channel,
which already on the BCS level depends on the $NN$ potential
\cite{Takatsuka93,Baldo98,Khodel98,Maurizio14,Srinivas16,Drischler17},
as at high density there is no constraint by the $NN$ phase shifts.
Furthermore TBF act generally attractive in this channel,
but effective mass and quasiparticle strength reduce the gap,
and polarization effects on $V$ might be of both signs,
in particular in asymmetric beta-stable matter
\cite{
Khodel04,            %POL,+,>1.
Zhou04,              %TBF,ANM,+,1.0
Schwenk04,           %POL,-,10^-4
Dong13,              %TBF+Z,-,0.1
Ding16,              %SCGF,-,NOPOL,0.1
Drischler17,         %TBF,+,0.5
Papakonstantinou17,  %UIX,+-,?
Ramanan20,           %REV,FIG11
Krotscheck23,Krotscheck24}. %PARQUET,-,0
Note that most theoretical investigations so far
consider only pure neutron matter.
Thus, due to the high-density nature of this pairing,
the various medium effects might be very strong and competing,
and there is still no reliable quantitative theoretical prediction for this gap.
The most recent investigation \cite{Krotscheck23,Krotscheck24}
points to a complete disappearance of the gap,
but previous works predict enhancement instead
\cite{Khodel04,Zhou04,Drischler17}.

Further important ingredients in the cooling simulations
are the neutron and proton effective masses,
which we actually used in \cite{Taranto16}.
In the BHF approach,
the effective masses can be expressed self-consistently
in terms of the s.p.~energy $e(k)$ \cite{Baldo14b},
\be
 \frac{m^*(k)}{m} = \frac{k}{m} \left[ \frac{d e(k)}{dk} \right]^{-1} \:.
\ee
We actually found \cite{Taranto16,Wei20} that
their effect can be absorbed into a rescaling of the 1S0 BCS pairing gap,
which we perform by introducing global scaling factors $s_y$ and $s_x$
on both magnitude and extension of the gap, i.e.,
\be
 \Delta(\rho) \equiv s_y \Delta_\text{BCS}(\rho/s_x) \:.
\label{e:sxy}
\ee
As in \cite{Wei20},
also in this paper we employ the same procedure,
and $s_y,s_x$ will be considered as free parameters in the cooling calculations.
Their optimal values will be determined later
in a combined analysis of cooling data and deduced NS mass distributions.

%Those gap functions are displayed in Fig.~\ref{f:gapp}
%as a function of particle ($n$ or $p$) density
%for different choices of $s_x=0.4,1.4(0.2)$ and $s_y=0.2,1.0(0.2)$, thus
%showing a reduction of both magnitude and
%density domain due to various in-medium effects
%which are simulated in a general way by this scaling procedure.
%Thus scaling factors larger than one appear very unrealistic.
%We include such choices only for the sake of a systematic investigation
%of their effect on the cooling.

To illustrate the important role played by the superfluidity gaps,
we first evaluate over which range of baryon density inside a NS
they are effective.
The starred markers in Fig.~\ref{f:MR} label the configurations $\m1s0$,
for which the BCS p1S0 gap vanishes in the NS center, i.e.,
$M<\m1s0$ stars are superfluid throughout,
whereas for $M>\m1s0$ there is a growing non-superfluid core region.
We see an important difference among the three EOS:
whereas for the V18 $\m1s0=1.92\ms$
and thus for heavier stars proton superfluidity is only partially present,
in the RMF cases the proton gap is fully active in all configurations.
Just as the absence of DU cooling,
also this feature is due the small proton fractions of the RMF models,
which causes relatively small proton partial densities.
Thus in this case NS cooling proceeds through the MU and BS processes,
modulated by superfluidity.

\begin{figure*}%...............................................................
\centerline{\includegraphics[width=1.15\textwidth]{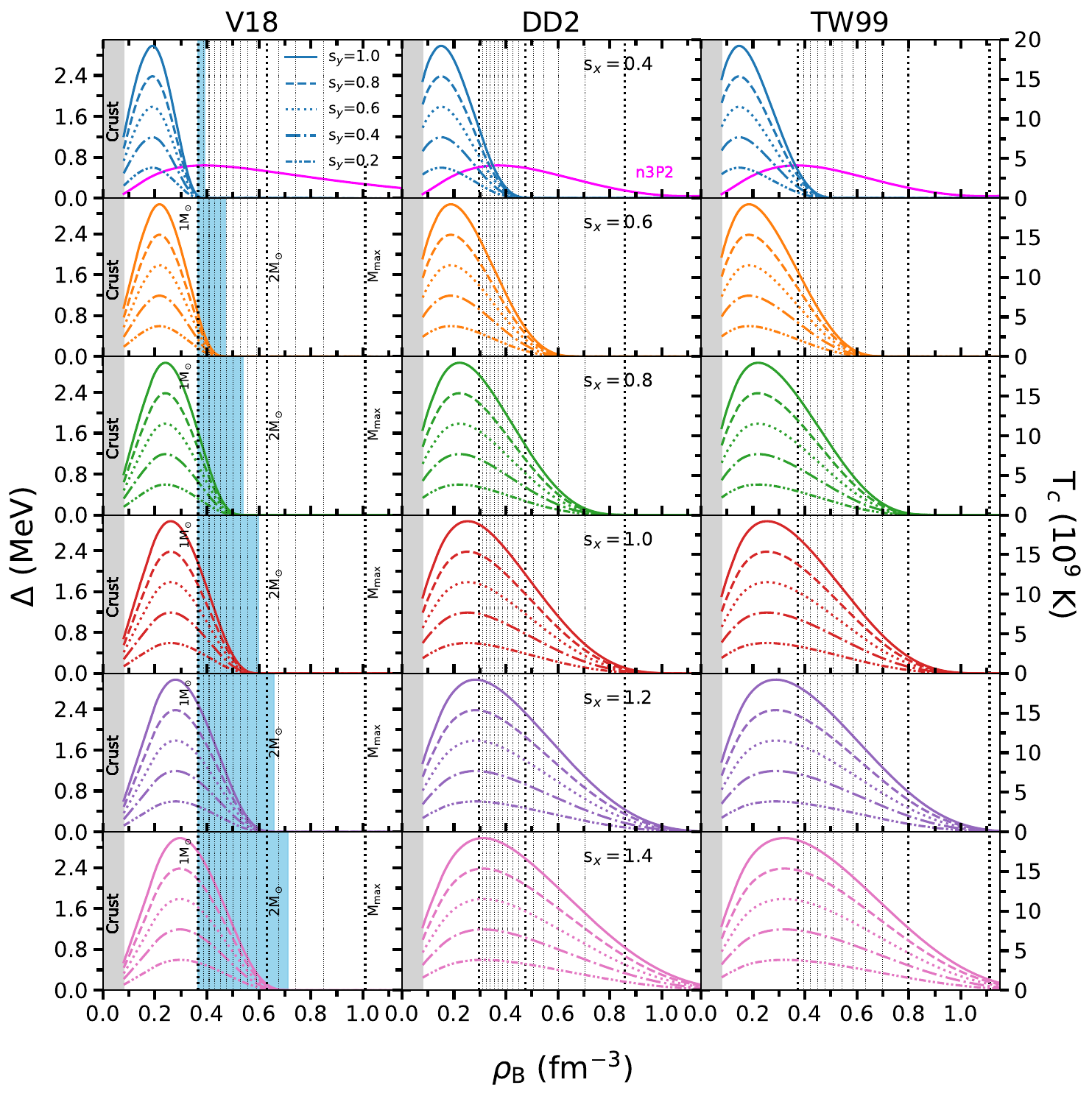}}
\vspace{-4mm}
\caption{
The BCS p1S0 pairing gap in NS matter
as a function of baryon density for the adopted EOSs,
with different scale factors $s_x$ (rows)
and $s_y$ (line styles) applied,
Eq.~(\ref{e:sxy}).
The shaded areas (only V18) indicate the density range
in which the DU process is blocked by superfluidity.
Vertical dotted lines indicate the central densities for
different NS masses $M/\!\ms=1.0,1.1,1.2,\ldots$ up to the maximum mass.
The unscaled n3P2 BCS gap is also shown in the upper panels.
\hj{}
}
\label{f:gap}
\end{figure*}

However, the parameter $s_x$ changes the active range of pairing,
and this is illustrated in Fig.~\ref{f:gap} for the chosen EOSs,
which shows the p1S0 gaps in NS matter
for several combinations of the scaling parameters $(s_x,s_y)$.
In each panel we indicate the central densities for several NS masses
(vertical lines) and the mass ranges (shaded regions, only V18)
in which DU cooling is suppressed by the p1S0 gap.
Due to the smaller proton fractions,
the gap extension over the baryon density range is larger
for both RMF models than the V18,
such that for $s_x\geq1$ pairing is always fully active in all stellar
configurations.
For the V18 on the contrary, for any choice of $s_x$,
there are always heavy NSs in which the DU process is unblocked,
causing very rapid cooling.
This is essential for the confrontation with cooling data
in the next section.

\begin{table}[t]%..............................................................
\caption{
NS matter baryon densities $\r1s0$
for the vanishing of the p1S0 gap
and corresponding NS masses $\m1s0$
with that central density,
as a function of the scale parameter $s_x$.}
%\medskip
%\def\myc#1{\multicolumn{1}{c}{$#1$}}
\renewcommand{\arraystretch}{1.05}
\setlength{\tabcolsep}{5pt}
\def\mr#1{\multirow{5}{*}{#1}}
\begin{ruledtabular}
\begin{tabular}{l|cccc}
%after \\: \hline or \cline{col1-col2} \cline{col3-col4} ...
EOS & $s_x$  & $\r1s0$ $[\!\fm3]$ & $\m1s0$ [$\!\ms$] \\
\hline
\mr{V18}
& 0.2 & 0.300 & 0.70 \\
& 0.4 & 0.388 & 1.11 \\
& 0.6 & 0.467 & 1.46 \\
& 0.8 & 0.536 & 1.73 \\
& 1.0 & 0.599 & 1.92 \\
& 1.2 & 0.657 & 2.06 \\
& 1.4 & 0.711 & 2.16 \\
\hline
\mr{DD2}
& 0.2 & 0.307 & 1.07 \\
& 0.4 & 0.495 & 2.06 \\
& 0.6 & 0.678 & 2.38 \\
& 0.8 & 0.861 & 2.44 \\
& 1.0 & 1.044 &      \\ %2.41
& 1.2 & 1.227 &      \\ %2.35
& 1.4 & 1.412 &      \\ %2.29
\hline
\mr{TW99}
& 0.2 & 0.315 & 0.75 \\
& 0.4 & 0.521 & 1.54 \\
& 0.6 & 0.710 & 1.91 \\
& 0.8 & 0.892 & 2.05 \\
& 1.0 & 1.073 & 2.09 \\
& 1.2 & 1.254 &      \\ %2.08
& 1.4 & 1.436 &      \\ %2.05
\end{tabular}
\end{ruledtabular}
\label{t:sx}
\end{table}%...................................................................

%.................................................................

In Table~\ref{t:sx} we summarize the values of the (central) density
at which the p1S0 gap disappears, $\r1s0$,
and the corresponding NS mass $\m1s0$,
for several values of the scaling factor $s_x$.
For completeness, we also display the unscaled n3P2 BCS gap
in the upper panels of Fig.~\ref{f:gap}.
It extends over the entire mass and density range,
and therefore would block all cooling processes for all NSs.
However, the competing n3P2 PBF process provides too strong cooling
for old objects, in disagreement with some data,
as found in \cite{Grigorian05,Wei20}
and confirmed in the following section.

\begin{figure*}%...............................................................
\vspace{-4mm}
\centerline{\includegraphics[width=1.15\textwidth]{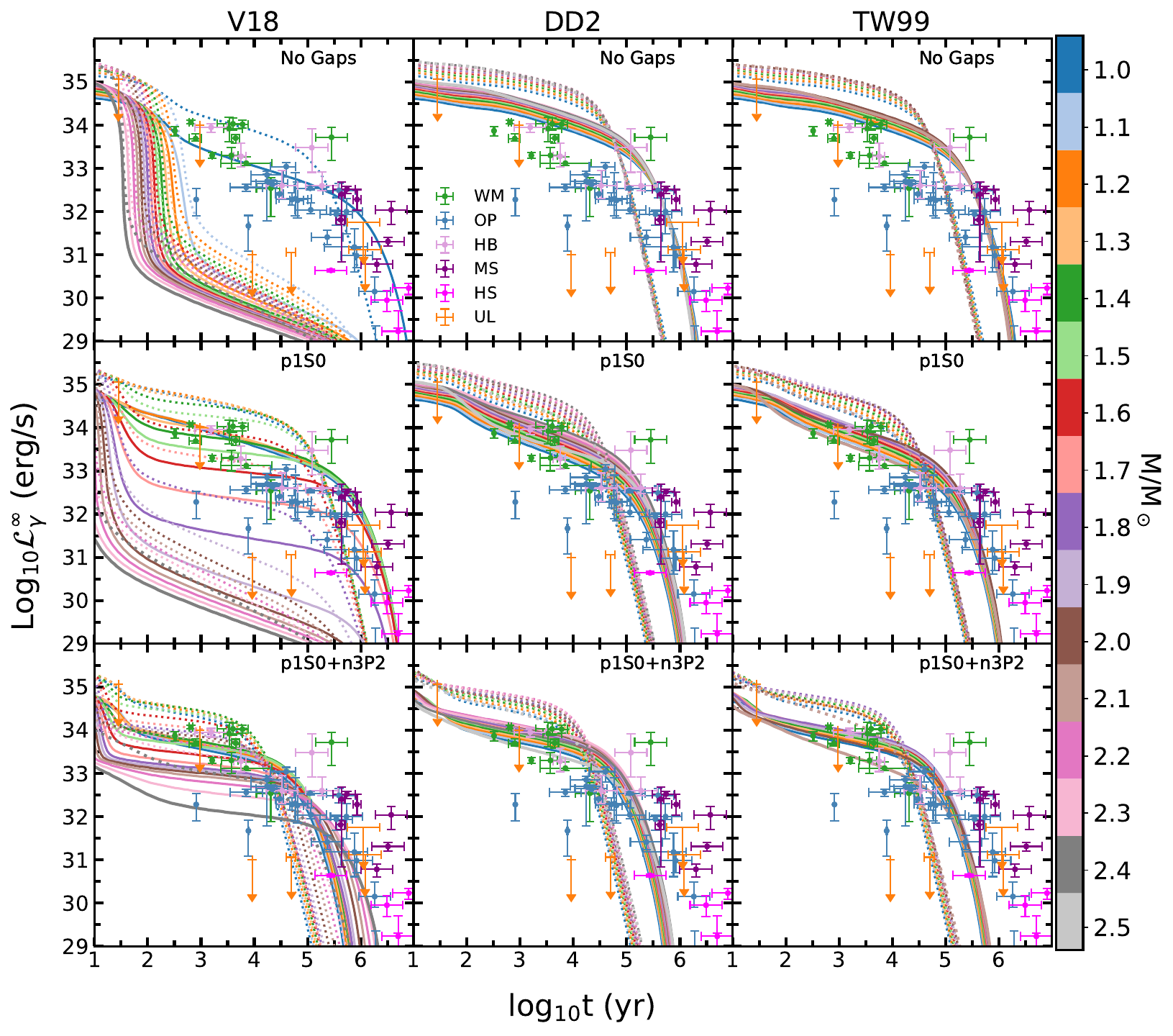}}
\vspace{-4mm}
\caption{
Cooling diagrams
obtained with different EOSs (columns)
without superfluidity gaps (upper row),
with p1S0 gap (middle row),
and with p1S0+n3P2 pairing gaps (lower row),
for different NS masses $M/\!\ms=1.0,1.1,\ldots,\mmax$
(decreasing curves).
The solid (dashed) curves are obtained with a Fe (light elements) atmosphere.
The data points are from \protect\cite{Potekhin20,Cooldat}.
See text for details.
}
\label{f:cool}
\end{figure*}%.................................................................

%==============================================================================
\section{Cooling simulations}
\label{s:res}

For the NS cooling simulations we employ the widely known
one-dimensional code {NSCool} \cite{Pageweb},
based on an implicit scheme developed in \cite{Henyey64} for solving
the general-relativistic equations of energy balance and energy transport,
\bal
 &\frac{1}{4\pi r^2}
 \sqrt{1-\frac{2Gm}{r}} e^{-2\phi} \frac{\partial}{\partial r}(e^{2\phi}L) =
 -Q_\nu - \frac{C_v}{e^\phi} \frac{\partial T}{\partial t} \:,
\\
 & \frac{L}{4\pi r^2} =
 -\kappa\sqrt{1-\frac{2Gm}{r}} e^{-\phi}\frac{\partial}{\partial r}(e^\phi T) \:.
\eal
Local luminosity $L$ and temperature $T$ depend on the
radial coordinate $r$ and time $t$.
The metric function is denoted by $\phi$, whereas
$Q_\nu$, $C_v$, and $\kappa$ are the neutrino emissivity,
the specific heat capacity, and the thermal conductivity, respectively.

The code solves the partial differential equations
on a grid of spherical shells.
The simulation is performed by artificially dividing
the star into two parts at an outer boundary
given by the radius $r_b$ and density $\rho_b=10^{10}\gc3$.
At $\rho<\rho_b$,
the so-called envelope \cite{Beznogov21}
includes the mass and composition change,
for instance, due to the accretion, and is solved separately in the code.
Here, we use the envelope model obtained in \cite{Potekhin97}.
At $\rho>\rho_b\ (r<r_b)$,
the matter is strongly degenerate
and thus the structure of the star is supposed to be unchanged with time.
As a result, we obtain for a NS of given mass and composition of the envelope,
a set of cooling curves showing the luminosity $L$ as a function of NS age $t$.
Each simulation starts with a constant initial temperature profile,
$\tilde{T}\equiv Te^{\phi}=10^{10}\,$K,
and ends when $\tilde{T}$ drops to $10^4\,$K.
Regarding the most important ingredient -- neutrino emissivity,
this code comprises all relevant cooling reactions:
nucleonic DU, MU, BS, and PBF,
including modifications due to p1S0 and n3P2 pairing.
Updates for the PBF rates \cite{Leinson10} are included.
Moreover, various processes in the crust are included,
such as the most important electron-nucleus bremsstrahlung,
plasmon decay, electron-ion bremsstrahlung, etc.

\begin{figure*}%...............................................................
\centerline{
\hspace{-8mm}\includegraphics[width=1.15\textwidth]{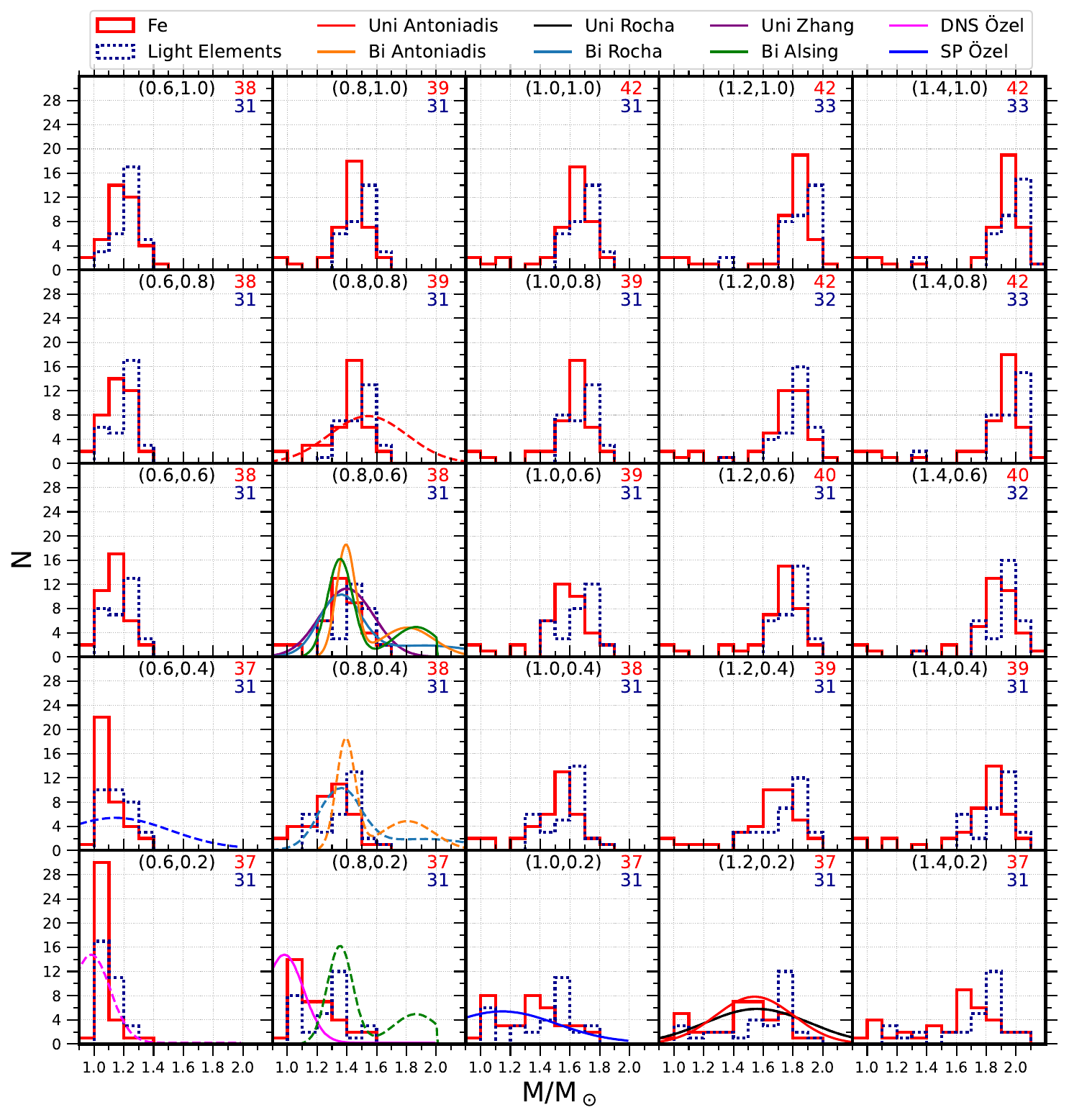}}
\vspace{-5mm}
\caption{
Deduced NS mass distributions for scaling factors
$(s_x=0.6,\ldots,1.4) \otimes (s_y=0.2,\ldots,1.0)$
and with Fe (solid red histograms)
or light-elements (dotted blue histograms) atmosphere.
$N$ is the number of data points lying in a given mass interval
$\Delta M=0.1\ms$
in the proper ($s_x,s_y$) cooling diagram.
Several panels show the best-fit theoretical results
\cite{Zhang11,Antoniadis16,Alsing18,Rocha19,Ozel12}
of Fig.~\ref{f:NStheo} superimposed,
using solid (dashed) lines for the Fe (light-elements) results.
The top-right numbers indicate the number of data points in the histograms.
}
\label{f:NSdist}
\end{figure*}%.................................................................

\begin{figure}[t]%.............................................................
\centerline{\includegraphics[width=0.5\textwidth]{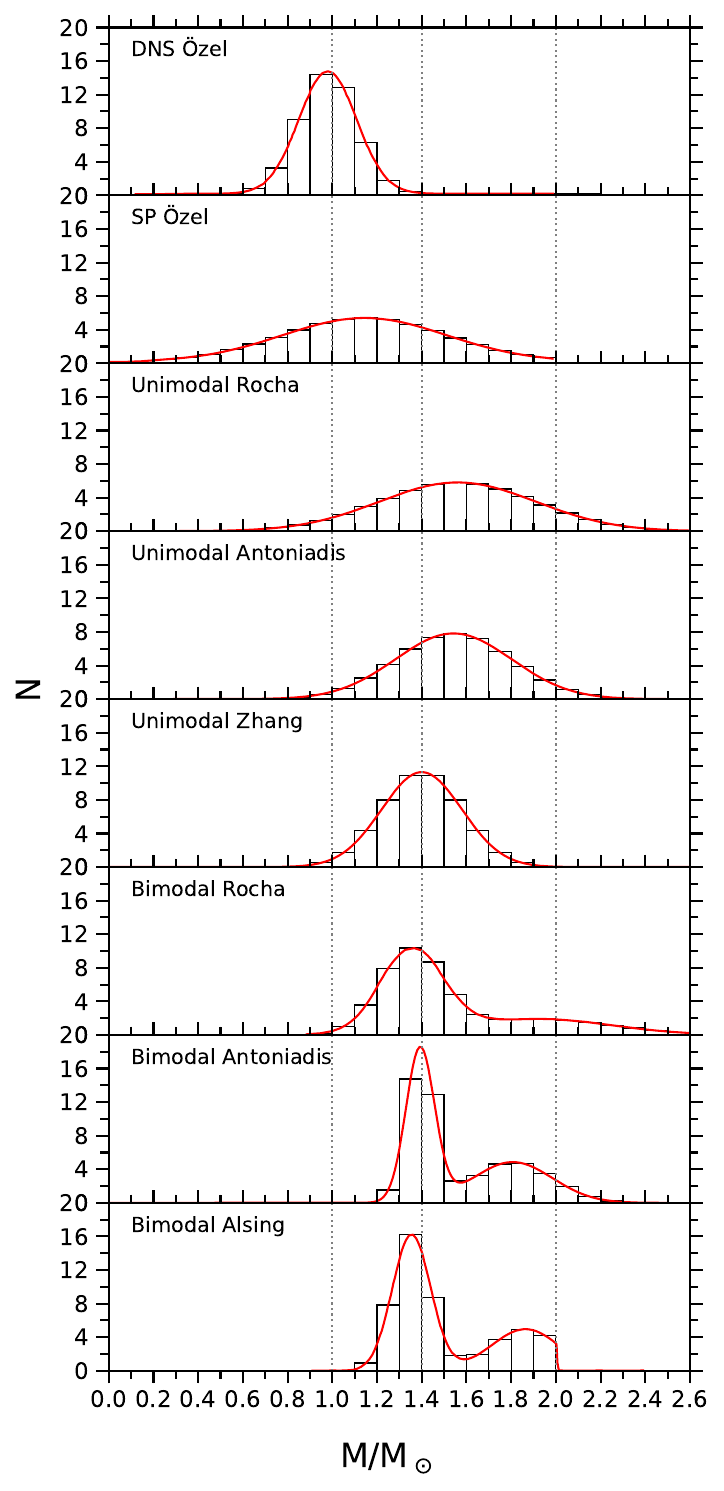}}
\vspace{-4mm}
\caption{
Theoretical NS mass distributions
\cite{Zhang11,Antoniadis16,Alsing18,Rocha19,Ozel12} (red curves)
binned in the same way as Fig.~\ref{f:NSdist} (black histograms).
The normalization is to 51 points.
Vertical lines indicate $M=1.0,1.4,2.0\ms$.
}
\label{f:NStheo}
\end{figure}%..................................................................

\begin{figure*}[t]%............................................................
%\centerline{\includegraphics[width=1.0\textwidth]{coolrmf8}}
\vspace{-4mm}
\centerline{\includegraphics[width=0.5\textwidth]{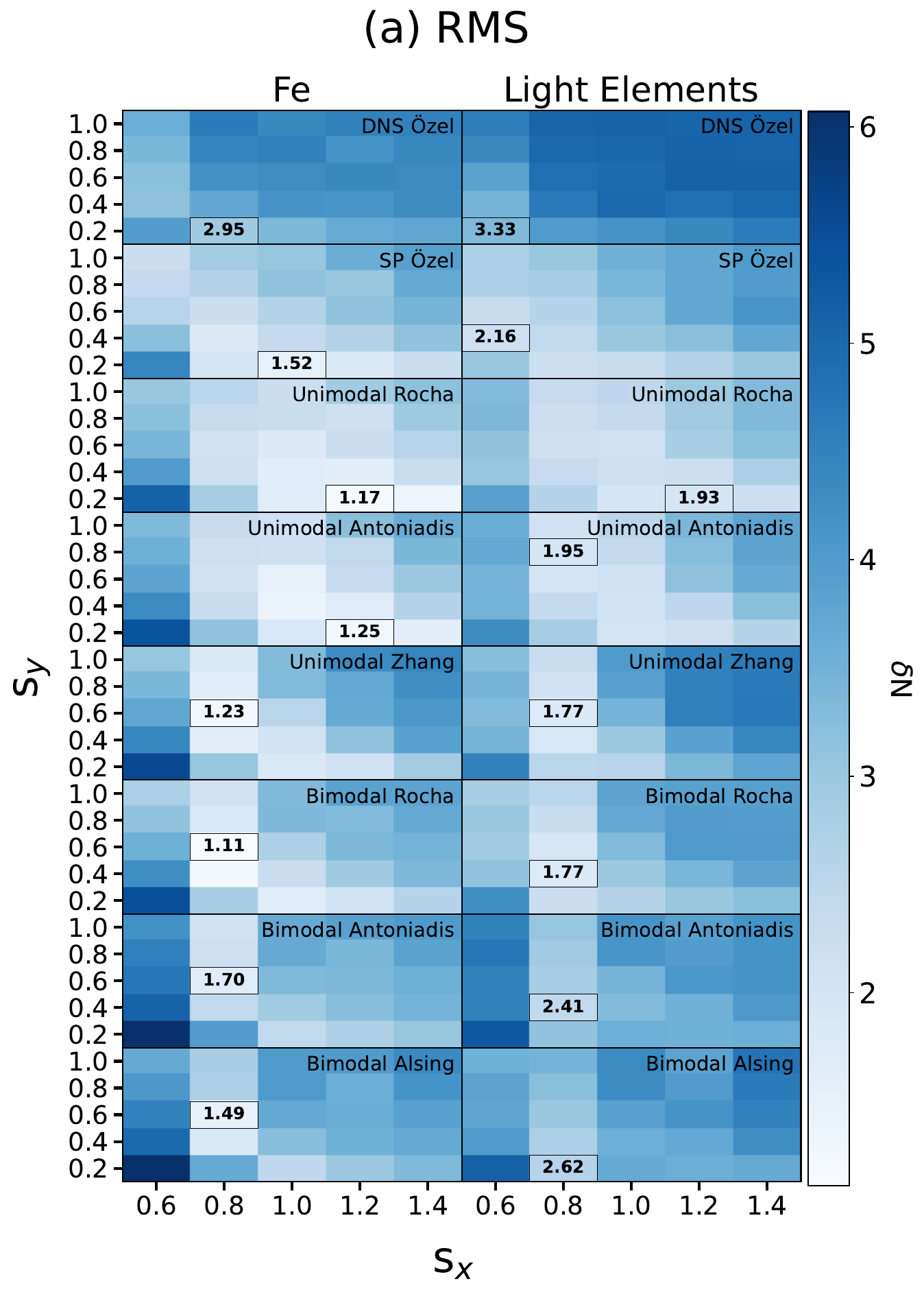}
\includegraphics[width=0.51\textwidth]{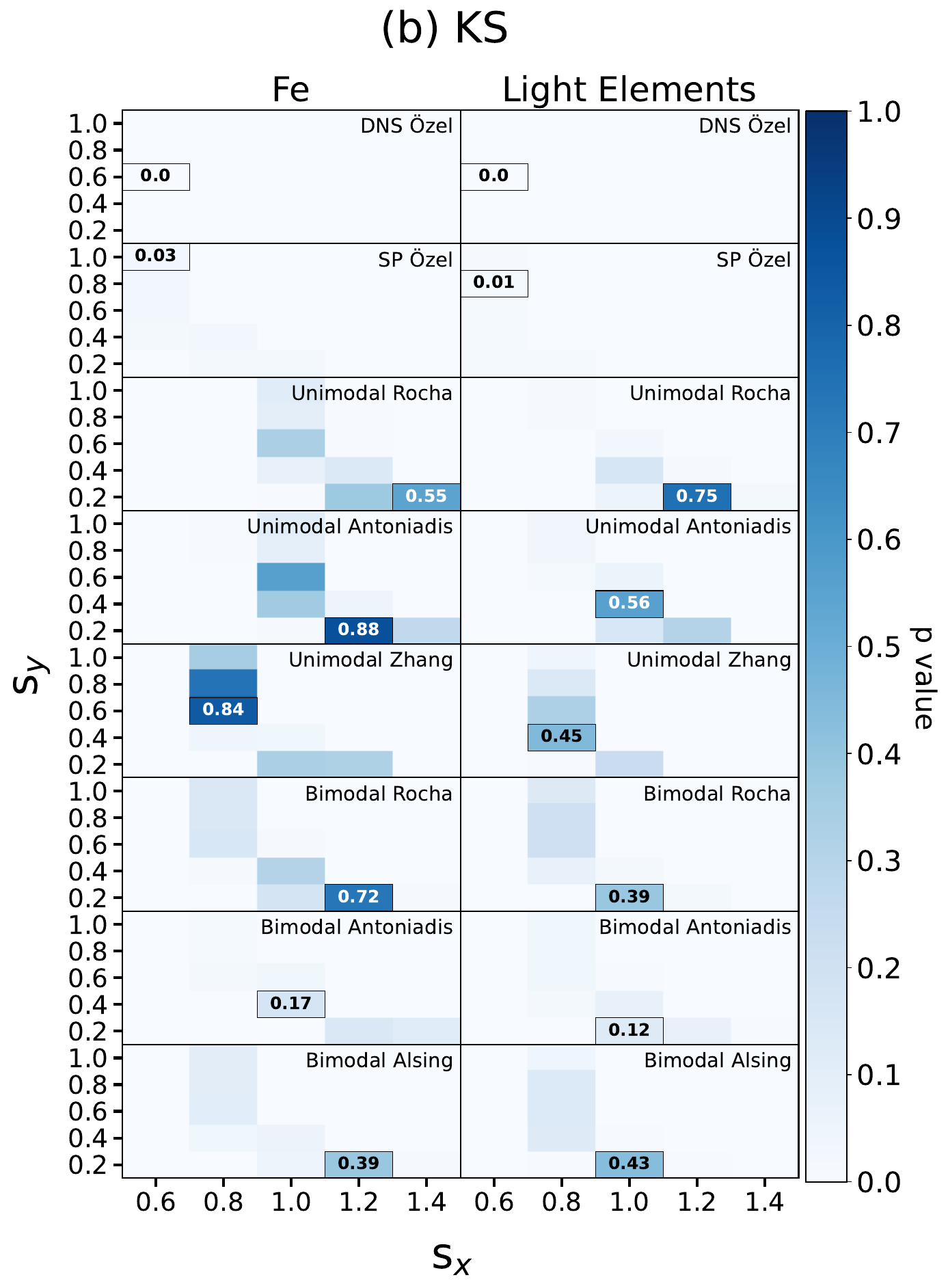}}
\vspace{-4mm}
\caption{
(a) Root-mean-square variance $\delta N$, Eq.~(\ref{e:rms}),
and (b) Kolmogorov-Smirnov $p$ value
between deduced NS mass distributions and various theoretical distributions
(shown in Fig.~\ref{f:NStheo}),
obtained with different p1S0 gap scale factors $s_x,s_y$
for a Fe or light-elements atmosphere.
Preferred values are indicated in squared boxes.
}
\label{f:heatmap}
\end{figure*}%.................................................................

In Fig.~\ref{f:cool} we show the cooling diagrams
obtained with the different EOSs,
employing a Fe atmosphere (solid curves)
or a light-elements atmosphere (dashed curves),
in comparison with the currently known data points
\cite{Beznogov15,Potekhin20,Cooldat}
(with partially large (estimated) error bars),
namely for
weakly-magnetized NSs (WM, green),
ordinary pulsars (OP, blue),
high-magnetic-field-B pulsars (HB, pink),
the magnificent seven (MS, violet),
small hot spots (HS, magenta), and
upper limits (UL, orange).
There are altogether 57 data points,
but we do not use the 6 UL data for the following analyses.

The upper row displays the results obtained without superfluidity.
One notes the strong effect of the DU process in the V18 EOS
characterized by a far too rapid cooling for all NSs with
$M>\mdu=1.01\ms$,
which is clearly unrealistic.
On the contrary the DD2 and TW99 EOSs without DU process
produce too slow cooling for middle-aged objects and too fast for old objects.
Thus the assumption of no superfluidity appears inconsistent
with both including or not the DU process,
regardless of the atmosphere model.

Accordingly in the middle row we include the p1S0 BCS ($s_x=s_y=1$) gap.
The main effect is the quenching of the DU process for the V18 EOS,
such that stars in the overlap zone
$1.01\ms = \mdu < M < \m1s0 = 1.92\ms$
cool moderately fast now,
and only high-mass stars, $M>1.92\ms$, cool very rapidly.
Altogether a very satisfying coverage of the data is achieved,
also taking advantage of the two atmosphere models.
In fact all data points can be covered within their error bars.
On the contrary, both no-DU models can still not explain most data.
%the old NS data.

In the lower row we investigate the effect of the n3P2 BCS gap:
This extends over the entire density range,
and therefore blocks DU processes for all NSs.
On the other hand the competing n3P2 PBF process provides a too strong cooling
for the old ($\gtrsim10^6$yr) objects \cite{Wei20b},
with or without DU process,
as already found by other authors
\cite{Grigorian05,Beznogov15,Beznogov15b,Beznogov18,Potekhin19,Potekhin20,Wei20}.
As a side remark,
also the proposed explanation of the claimed fast cooling of the Cas~A NS
by a suitably chosen n3P2 gap is problematic,
see \cite{Shternin23,Leinson22} for reference and recent works,
apart from the fact that the claim itself is disputed as being
due to detector degradation \cite{Posselt22}.

%We also remind the possible strong constraints on NS cooling imposed by the
%speculated very rapid cooling of the supernova remnant Cas~A
%\cite{2009Nat,2010HeiHo,2013Elsha,2015HoPRC},
%which we have studied in detail in \cite{Taranto16}.
%As the observational claims remain highly debated \cite{casno1,casno2},
%we do not consider this scenario in this work.
%\hj{best sx sy plot?}

In conclusion,
an EOS featuring DU cooling for a wide enough mass range of NSs
together with (partial) quenching by the p1S0 BCS gap
seems to be required to reproduce the cooling data.
It seems difficult to accommodate finite n3P2 pairing in this setup.
We therefore continue the analysis with the V18 EOS including p1S0 pairing
but without n3P2 pairing.
We have not investigated if a fine-tuned n3P2 gap can still be adopted
within our scenario,
but it is not required for reproducing current data.

%We remark that the approach presented here
%is able to describe perfectly with the same microphysics input
%not only the cooling of isolated NSs discussed here,
%but also the cooling of
%reheated accreting NSs in X-ray transients in quiescence (XRTQ)
%\cite{Yakovlev14,Beznogov15,Beznogov15b},
%as shown in \cite{Fortin18}.

%==============================================================================
\section{Gaps and mass distributions}
\label{s:cor}

Currently no information on the actual masses of the various cooling objects
is available,
hence a comparison between theoretically predicted masses
and the actual masses of the cooling diagram is not possible.
In this situation we can {\em derive} a NS mass distribution
that is consistent with the outcome of a given cooling simulation
by simply counting the number of data points that fall into a given
interval between two adjacent fixed-$M$ cooling curves
\cite{Wei19,Wei20}.
(Error bars are disregarded at this level of investigation).
The resulting histogram can be compared with mass distributions of NSs
obtained in different, independent theoretical ways
\cite{Zhang11,Ozel12,Kizil13,Antoniadis16,Alsing18,Rocha19,Landry21}.

In doing so,
we assume that the mass distribution of isolated NSs in the cooling diagram is
not different from the mass distributions of NSs in binary systems
\cite{Zhang11,Antoniadis16,Alsing18,Rocha19,Farrow19}
or all NSs in the Universe,
and that the detection of these sources is independent of their brightness
(Malmquist bias \cite{Wall12});
both of these assumptions being highly unlikely to be fulfilled,
and we will estimate their importance later.
A further principal problem is the lack of information on the atmosphere
of the data objects,
which requires further theoretical assumptions in this analysis,
the simplest one being to use a fixed atmosphere model.

In attendance of future improvement of information on the data sources,
we proceed nevertheless with this way of derivation of the mass distribution:
The masses of the 51 cooling data points are assumed to be those predicted
theoretically by their position in the cooling diagrams
among the theoretical curves displayed in Fig.~\ref{f:cool},
and Fig.~\ref{f:NSdist} shows the resulting mass histograms for different
choices of the p1S0 pairing parameters ($s_x,s_y$),
assuming either a common Fe (solid histograms)
or a light-elements atmosphere (dotted histograms)
for all data points.
This is clearly unrealistic,
and in fact in the first case up to 42 and in the second case only up to 33
%[apart from 10 for (0.6,0.2) and 12 for (1.4,0.8) and (1.4,1.0)]
of the 51 sources can be fitted,
while a fit of more data would require a suitable choice of atmosphere
for each object.
In this case only 4 objects would remain out of this analysis
for the best parameter choices
(but their errors bars still reach the external theoretical curves):
Calvera, J0806, J2143, J1154.

One observes that increasing $s_x$ or to a lesser degree $s_y$
shifts the centroid of the derived mass distributions to higher values,
since the cooling curves move upwards due to the increased suppression
of the DU process,
as shown in Fig.~4 of \cite{Wei20}.

The different results in Fig.~\ref{f:NSdist} can now be compared with
other theoretical studies of the NS mass distribution
\cite{Zhang11,Ozel12,Kizil13,Antoniadis16,Alsing18,Rocha19},
in particular unimodal or bimodal distributions,
which were derived on the basis of distinct evolutionary paths
and accretion episodes.
In this paper we also include the analysis \cite{Ozel12}
for double neutron stars (DNS) and slow pulsars (SP).
Fig.~\ref{f:NStheo} contains a compilation of the theoretical mass distributions
that we use for comparison.
We provide a binning and normalization consistent with Fig.~\ref{f:NSdist}
in order to confront directly with our results.
For a quantitative comparison we simply compute the rms deviations between
the histograms
$\{N_i^\text{dat}\}$ in Fig.~\ref{f:NSdist} and
$\{N_i^\text{theo}\}$ in Fig.~\ref{f:NStheo},
\be
 \delta N \equiv \sqrt{ \frac{1}{N^\text{dat}}
 \sum_i \Big(N_i^\text{dat} - N_i^\text{theo}\Big)^2 }  \:,
\label{e:rms}
\ee
where $i$ labels the mass bins and $N^\text{dat}=51$
is the total number of data points excluding the UL markers.

The results are visualized in the heatmap shown in Fig.~\ref{f:heatmap}(a)
for the various combinations,
where we also indicate the optimal values $s_x,s_y$
(those predicting the smallest $\delta N$)
for each theoretical mass distribution
and the two atmosphere models separately.
Of course the use of a unique atmosphere model for the whole data set
is unrealistic,
but currently impossible to overcome without further constraints on the data.
Nevertheless some qualitative conclusions can be drawn:

For a Fe atmosphere
the best agreement with most considered distributions
seems to require values of $s_x\approx0.8$ and $s_y\approx0.6$,
which would also be consistent with microscopic investigations
of the 1S0 pairing gap, as discussed before.
The quality of agreement is worst for the DNS \"Ozel one.
%with a mass median of about $1\ms$.
%which favors a very large $s_x$.
In Fig.~\ref{f:NSdist} [panels (0.6,0.2) and (0.8,0.2)]
it can clearly be seen that the mass median $\sim1\ms$
of this distribution is
unrealistically low.
%far too low compared to the data.
The best overall fit is provided by the
Bimodal Rocha distribution
(median $\sim1.4\ms$)
and $(0.8,0.6)$.
The corresponding panel in Fig.~\ref{f:NSdist}
shows a decent agreement between solid histogram
and corresponding theoretical curves,
confirming the numerical analysis.
Also Unimodal Rocha, Unimodal Antoniadis,
and Unimodal Zhang distributions
provide fits of similar quality.
All these distributions have in common that their main peak
is located at about $1.4\ms$.
The existence of a minor second peak
or in fact the extended tail of the Bimodal Rocha distribution
cannot be rejected clearly by the current data.

It is remarkable that
also for a light-elements atmosphere,
the theoretical models Bimodal Rocha and Unimodal Zhang
and $s_x=0.8$ and $s_y=0.4$--$0.6$ are singled out as preferred values,
but the quality of the fits is worse than for the Fe atmosphere,
which is also due to the fact that
less data points are comprised in the analysis (31--33)
than for the Fe atmosphere (37--42).

The ($s_x,s_y$) configurations which fit best a given theoretical model,
have that theoretical curve superimposed in Fig.~\ref{f:NSdist},
and five of the models
single out (0.8,0.6) as `best' parameter set.

\begin{figure}[t]%............................................................
\centerline{\includegraphics[scale=0.65]{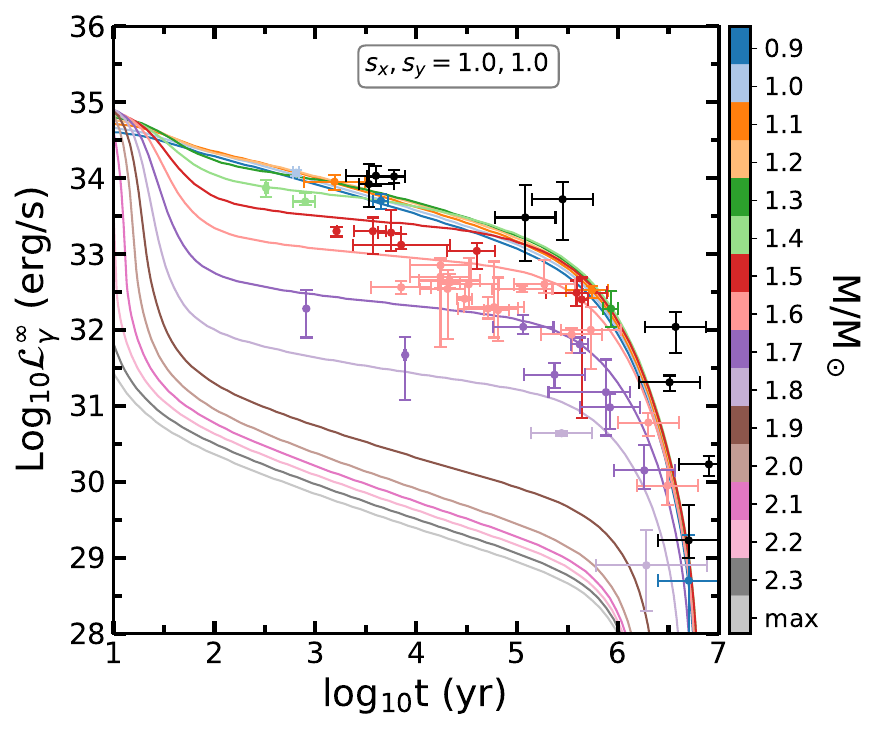}}
\vspace{-4mm}
\caption{
Mass assignment to the various data points
for $(s_x,s_y)=(1,1)$ and Fe atmosphere,
using the same color as the closest upper fixed-$M$ cooling curve.
Black data are not in the mass histogram.
}
\label{f:datm}
\end{figure}%.................................................................

\begin{figure}[t]%............................................................
\centerline{\includegraphics[scale=0.6]{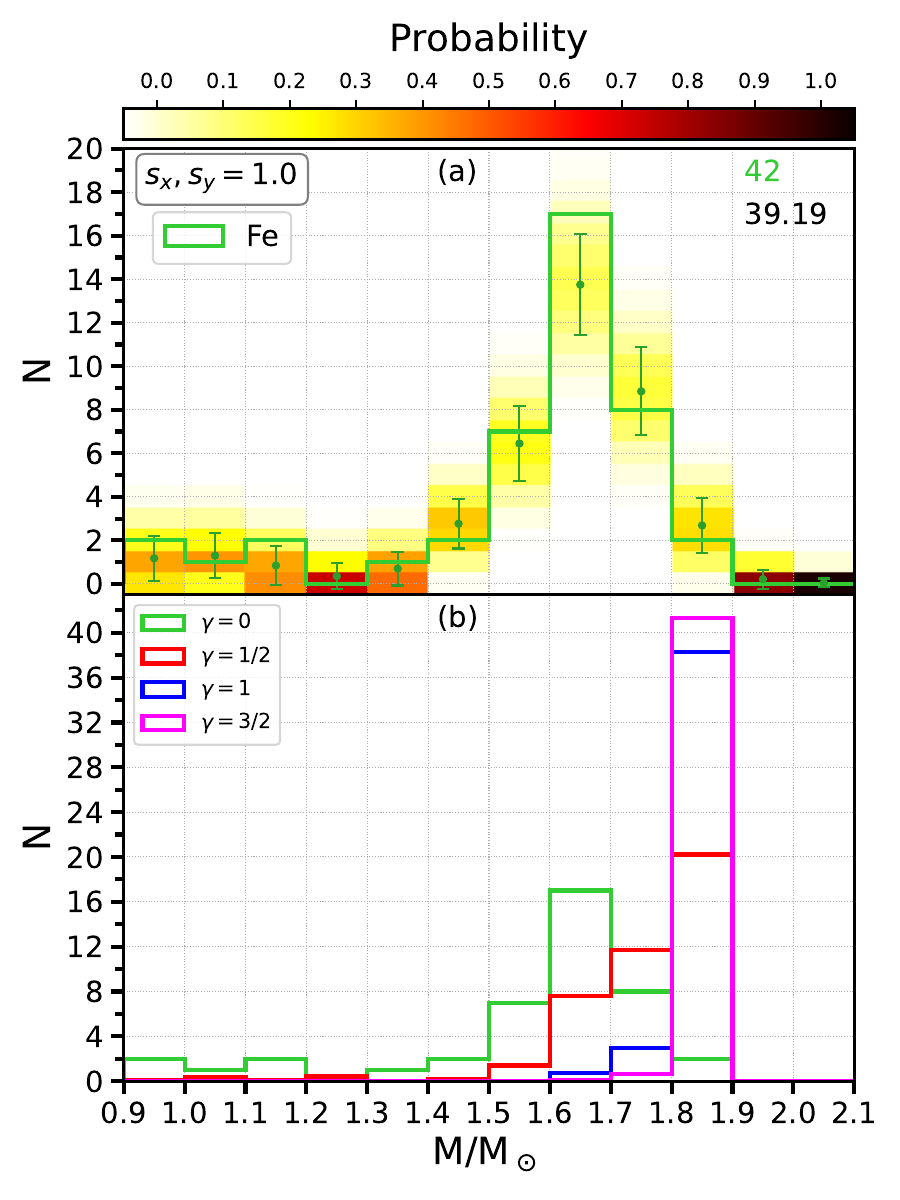}}
\vspace{-4mm}
\caption{
(a)
Mass-distribution histogram
for $s_x=s_y=1$ and a Fe atmosphere,
smoothed by inclusion of the data error bars in a Monte-Carlo method,
see text for details.
The mean values and variances for each mass bin are indicated by error bars.
The solid green histogram indicates the results disregarding data errors.
(b)
The effect of a luminosity correction $\sim L^{-\gamma}$
on the mass histogram.
}
\label{f:errors}
\end{figure}%.................................................................

We conclude that in most cases
the comparison between cooling diagrams and population models
indicates a reduction of the pairing range to $s_x\approx0.8$
and also a reduction of the magnitude $s_y\approx0.6$,
which is however less well determined and more model dependent.
This agrees qualitatively well with theoretical estimates of
the reduction of the BCS gap by medium effects \cite{Burgio21}.
While pronounced bimodal distributions appear slightly disfavored,
the current limitations of theoretical method and available data
do not allow to clearly identify a preferred theoretical population model,
though.
In particular, we stress again a probable selection effect that
would make heavy and faint NSs unobservable
and thus not appear in the high-mass part of the histograms.
This would put a bias on a too low centroid of the derived mass distributions
and accordingly too strong suppression of the p1S0 gap.
We illustrate this problem in the next section,
but it is currently impossible to quantify this assertion,
and our analysis remains at an exploratory state.

We have also carried out an alternative comparison
of derived and theoretical mass distributions
by a Kolmogorov-Smirnov test \cite{KS}.
This test is much more selective,
as demonstrated by the $p$ values
(consistency probability)
shown in Fig.~\ref{f:heatmap}(b).
Interestingly, Unimodal Zhang and Unimodal Antoniades are again
the best-fit distributions with the same preferred $(s_x,s_y)$ values
as for the RMS test,
while DNS \"Ozel, SP \"Ozel, and Bimodal Antoniades can be discarded
based on the KS test.

We conclude this section by showing in Fig.~\ref{f:datm}
the mass assignment to the various data points
for $(s_x,s_y)=(1,1)$ and Fe atmosphere.
It is obvious that due to the tight spacing between some adjacent
fixed-$M$ cooling curves,
the error bars of the data become significant.
Also this issue is discussed in the following section in more detail.

%==============================================================================
\section{Analysis of errors}
\label{s:err}

We investigate here briefly the main errors affecting the current procedure,
namely
(a) the statistical errors characterizing the current data, and
(b) the systematic selection effect favoring the detection and presence
of luminous sources in the data catalog.

Regarding (a), the fairly large error bars on both luminosity and age
of the various data sources in Fig.~\ref{f:cool}
imply that also the derived mass histograms in Fig.~\ref{f:NSdist}
should have associated errors.
We simulate them by a Monte-Carlo procedure generating a large number
of pseudo-data sets by assuming Gaussian variations of each data point
according to its associated error bars.
For each pseudo-data set a new histogram is generated
and those are accumulated and averaged.
A typical result it shown in Fig.~\ref{f:errors}(a)
in the form of a heat plot for the $s_x=s_y=1$, Fe-atmosphere histogram,
in comparison with the original bare histogram employing the original data set,
shown in the relevant panel of Fig.~\ref{f:NSdist}.

One can see that the bare histogram lies still within the error bars of the
averaged result and the centroid is not modified.
We conclude that even taking account of the substantial observational
error bars,
the deduced mass histograms are still fairly well defined to perform
the comparison with theoretical mass distributions as carried out in
this work.
However, we do not perform a complete error analysis of this aspect,
which seem futile facing the other systematic uncertainties
regarding data and theoretical mass distributions at the present stage,
in particular the following.

(b) the elimination of the bias on bright sources in a data sample
(Malmquist bias) is a very difficult problem \cite{Wall12}
that we do not attempt to solve here.
We rather illustrate the qualitative consequences of such a procedure
in the following.
For that we apply a luminosity correction factor $\sim L^{-\gamma}$
to each data point when filling the mass histograms
(and renormalize to the original histogram at the end).
In the naive Newtonian case $\gamma=3/2$,
which compensates for the observationally available volume,
but this is invalidated by general relativity
on cosmological scales \cite{Wall12}
or due to nonuniform distribution of NSs in our galaxy \cite{Popov03}.
We therefore choose empirically $\gamma=0,1/2,1,3/2$,
ranging from no correction (results used in this work)
to the exaggerated Newtonian correction.
The results are shown in Fig.~\ref{f:errors}(b),
again for the $s_x=s_y=1$, Fe-atmosphere case.

As expected, an increasing luminosity correction causes a shift of the
mass histogram to higher values,
as now the importance of massive (and therefore faint)
stars is amplified in the sampling.
However, the procedure cannot fill bins with masses higher than those
present in the original sample,
$M<1.9\ms$ in this case,
and is thus of limited use for too small samples:
Already for $\gamma=1/2$ the peak of the mass distribution
is located at its upper border.
Also, the most realistic way to remove the luminosity bias is not known,
and would require a much more profound data analysis than attempted
in this work,
apart from a better and much larger data set.
We thus leave such attempt for the future,
when at least a bigger data set will be available.

%==============================================================================
\section{Conclusions}
\label{s:end}

We have presented a global combined analysis of NS cooling and mass distributions,
employing the most recent set of cooling data.
We confirm that those data demand a fast cooling mechanism like the DU process
(or alternative scenarios
\cite{Grigorian05,Blaschke07,Blaschke13,Shternin18,Suleiman23}
not considered in this work).
This imposes an important constraint on the proton fraction and symmetry energy
of a realistic nucleonic NS EOS,
complementary to those obtained from nuclear structure,
global NS observables, and recent gravitational wave observations.
We employed here the V18 BHF EOS that fulfills all these constraints
and predicts DU cooling for even the lightest NSs.

Cooling can then be modulated by the strength of the p1S0 pairing gap,
while the PBF process associated with
n3P2 pairing seems to provide too strong cooling for a satisfactory
description of the data.
The best simultaneous reproduction of various theoretical NS mass distributions
is possible with a p1S0 gap slightly reduced in magnitude and density extension,
in qualitative agreement with theoretical investigations of that gap,
while the disappearance of the n3P2 gap is also supported by recent
theoretical investigations.

Currently this method is mainly hampered by missing information
on masses and atmospheres of the cooling data,
which will constitute very effective constraints in the future.
Also both statistical and in particular the systematic error associated with
Malmquist luminosity bias of the sources
are too important to allow quantitative conclusions,
so that our method should be considered as a proof-of-concept
that will become quantitative
when more abundant and precise data become available in the future.
This will also allow to pin down
the density range (onset density) of (blocked) DU cooling in the EOS,
which is another important degree of freedom
that was not studied in this work.

Our general framework was a purely nucleonic scenario,
and exotic types of matter like hyperons or quark matter
were not considered in this work.
Those still present a formidable challenge to cooling calculations
due to their largely unknown microphysics ingredients relevant for cooling,
like cooling rates, heat capacities, transport properties,
and most of all superfluid properties.

%==============================================================================
%\vspace{14mm}

\acknowledgments{
This work was partially funded by
the National Natural Science Foundation of China under Grant No.~12205260.
%and the China Scholarship Council, No.~201706410092.
%We acknowledge helpful communication with D.~Page.
}

%\appendix
%\section{}

%==============================================================================
%http://www.issn.org/services/online-services/access-to-the-ltwa/
%\newcommand\mathplus{+}
%\newcommand\mdash{--}
\newcommand{\araa}{Annu. Rev. Astron. Astrophys.}
\newcommand{\aap}{Astron. Astrophys.}
\newcommand{\apjl}{Astrophys. J. Lett.}
\newcommand{\epja}{EPJA}
\def\jpg{J. Phys. G}
\newcommand{\mnras}{Mon. Not. R. Astron. Soc.}
\def\npa{Nucl. Phys. A}
\newcommand{\physrep}{Phys. Rep.}
\newcommand{\plb}{Phys. Lett. B}
\def\ppnp{Prog. Part. Nucl. Phys.}
\newcommand{\ssr}{Space Science Reviews}

%\reftitle{References}
%\externalbibliography{yes}
\bibliography{coolrmf}

\end{document}